\documentstyle[aps]{revtex}

\newcommand{\pol}{\hat{\bf e}}
\newcommand{\rv}{{\bf r}}
\newcommand{\Ev}{{\bf E}}
\newcommand{\Dv}{{\bf D}}
\newcommand{\Pv}{{\bf P}}

\newcommand{\dv}{{\bf d}}

\newcommand{\eo}{\epsilon_0}
\newcommand{\beq}{\begin{equation}}
\newcommand{\eeq}{\end{equation}}
\newcommand{\bea}{\begin{eqnarray}}
\newcommand{\eea}{\end{eqnarray}}

\begin{document}
\draft
\preprint{}
\title{Lorentz-Lorenz shift in a Bose-Einstein condensate}
\author{Janne Ruostekoski$^1$ and Juha Javanainen$^2$}
\address{
$^1$Department of Physics, University of Auckland, Private Bag 92019,\\
Auckland, New Zealand\\
$^2$Department of Physics, University of Connecticut, Storrs,
Connecticut 06269-3046}
\maketitle
\begin{abstract}
We study the quantum field theory of light-matter interactions for
quantum degenerate atomic gases at low light  intensity. We argue that
the contact interactions between atoms emerging in the dipole gauge
may be ignored. Specifically, they are canceled by concurrent infinite
level shifts of the atoms. Our development yields the classic
Lorentz-Lorenz local-field shift of the atomic resonance. \end{abstract}

The understanding of the interactions of light with matter in quantum
degenerate
systems has become especially topical after the first evidence for a
Bose-Einstein
condensate (BEC) of an atomic gas has appeared \cite{bec}.
Recently, Andrews {\it et al}.\ \cite{AND96} have also reported on
non-destructive optical detection of a Bose condensate.

In the limit of large detuning of the driving light from atomic resonance, the
dynamics of the light and matter fields may be decoupled. It then turns out
that
the spectrum of the scattered light conveys direct signatures of atom
statistics
\cite{JAV95,GRA96}, and that under various conditions even the phase of
the macroscopic wave function may be observed optically
\cite{optdet}. On the other hand, for a dense enough sample
and small enough atom-field detuning, the analysis of the response of matter
requires a concurrent treatment of light with its own dynamics. The result is
that nearby atoms alter the optical response of each other. A large body
of work in this direction \cite{SVI90,POL91,JAV94,YOU95,YOU96,MOR95,RUO97a}
seems to have brought about the general notion that atom statistics should have
only a minor effect on absorption, dispersion or diffraction of light
\cite{JAV95,KAS96}. Nonetheless, even for the Maxwell-Boltzmann gas and for
the simplest optical properties such as refractive index, there still are no
proven solutions for microscopic theories regarding a dense, near-resonance
sample.

In this paper we continue our rigorous quantum field theoretical
analysis of light-matter interactions \cite{JAV95,RUO97a}. We argue that
the contact interaction terms between different atoms that arise in
the {\it length} gauge do not have any effect on light-matter dynamics.
As a result, for a model BEC we find precisely the Lorentz-Lorenz (LL)
shift familiar from classical electrodynamics
\cite{LOR,BOR75,JAC75}.

In the many-particle formalism of light-matter interactions, the
Hamiltonian is often transformed into the length gauge using
the Power-Zienau-Woolley transformation \cite{COH89}. The electric
displacement is then the basic dynamical degree of freedom, instead of
the electric field, and the Hamiltonian picks up a polarization
self-energy term
\beq
H_{\rm P}={1\over 2\eo}\int d^3 r\,\Pv(\rv)\cdot\Pv(\rv)=
{1\over2\eo}\,\sum_{i\neq j}
\dv_i\cdot\dv_j\,\delta(\rv_i-\rv_j)\,,
\label{eq:pse}
\eeq
where $\Pv(\rv)=\sum_{i}\dv_i\,\delta(\rv_i-\rv)$ is the electric
polarization in the dipole approximation and $\dv_i$ and $\rv_i$ are the
dipole operator and the center-of-mass (c.m.) position operator for the
$i^{\rm th}$ atom. On the right hand side of Eq.~{(\ref{eq:pse})} we have
ignored the divergent self-energies with $i=j$. Polarization self-energy 
describes a contact interaction between polarized atoms. One might argue that 
for a quantum degenerate Bose gas this term is important, as the de Broglie
wavelengths of the atoms are of the order of the interatomic distances
and the wave functions of different atoms overlap strongly. The
polarization energy has been duly included in many theoretical
approaches to light-atom interactions in the quantum degenerate regime
\cite{YOU95,MOR95,RUO97a,LEN94}.

In our previous paper \cite{RUO97a} we have derived a hierarchy of equations of
motion for correlation functions that contain one excited-atom field and
one, three, five, etc., ground state atom fields, for the limit of low
light intensity. All contact interactions, such as those produced by the
polarization self-energy (\ref{eq:pse}), were included. For the total
electric field $\Ev^+$ we found the following monochromatic expression
in the presence of scattering by atomic dipole moments
\beq
\eo{\bf E}^+({\bf r}) = {\bf D}^+_F({\bf r}) +
{1\over i\kappa}\int d^3r'\,
{\sf G}({\bf r}-{\bf r'})\,{\bf P}^+({\bf r}')\,,
\label{eq:MonoD}
\eeq
where ${\bf D}^+_F$ is the positive frequency component of the driving
electric displacement, the scalar constant $\kappa$ is defined in terms of the
reduced dipole moment ${\cal D}$ as $\kappa = {\cal D}^2/\hbar\epsilon_0$, and
the positive frequency part of the polarization operator ${\bf P}^+$ is
given in
second quantization by
\begin{equation}
{\bf P}^+({\bf r}) = {\bf d}_{ge}\psi^\dagger_{g}({\bf
r})\psi_{e}({\bf r})\,.
\label{eq:PDF}
\end{equation}
Here $\psi_g$ and $\psi_e$ are the ground state and the excited state
field operators in the Heisenberg picture, and $\dv_{ge}$ is the dipole
matrix element for the transition $g\rightarrow e$. 
In our notation the same level index appearing twice in a product indicates
a summation over the magnetic substates of the level. The monochromatic
version of the  $3\times3$ tensor propagator
${\sf G}({\bf r})$ is
\begin{equation}
{\sf G}_{ij}({\bf r}) = i\kappa\left\{
\left[ {\partial\over\partial r_i}{\partial\over\partial r_j} -
\delta_{ij} {\bbox \nabla}^2\right] {e^{ikr}\over4\pi r}
-\delta_{ij}\delta({\bf r})
\right\}\,.
\label{eq:GDF}
\end{equation}

The expression ${\sf G}({\bf r}-\rv'){\bbox {\cal D}}$ is equal
to the positive-frequency component of the electric field from a
monochromatic dipole with the complex amplitude ${\bbox {\cal D}}$,
given that the dipole resides at $\rv'$ and the field is observed
at $\rv$ \cite{JAC75}. The explicit expression is
\begin{equation}
{\sf G}({\bf r}){\bbox {\cal D}} =
{i\kappa\over4\pi}
\bigl\{ k^2(\hat{\bf n}\!\times\!{\bbox {\cal D}}
)\!\times\!\hat{\bf n}{e^{ikr}\over r} +[3\hat{\bf n}(\hat{\bf
n}\cdot{\bbox {\cal D}})-{\bbox {\cal D}}]
\bigl( {1\over r^3} - {ik\over r^2}\bigr) e^{ikr}
\bigr\}-{i\kappa\over3}\,{\bbox {\cal D}}\,\delta({\bf r})\,,
\label{eq:DOL}
\end{equation}
with $\hat{\bf n} = {{\bf r}/ r}$ and $k = {\Omega / c}$, $\Omega$
being the dominant frequency of the incident light field. The volume
integral over $1/r^3$ in Eq.~{(\ref{eq:DOL})} is not absolutely
convergent in the neighborhood of the origin. The expression
{(\ref{eq:DOL})} should be understood in such a way that the integral of
the term inside the braces over an infinitesimal volume enclosing the
origin vanishes \cite{RUO97a}. Equation {(\ref{eq:DOL})} is precisely the
classical expression of the dipolar field, including the peculiar delta
function divergence  at the origin \cite{JAC75}.

With the definitions
\begin{mathletters}
\bea
{\bf P}_l({\bf r}_1,\cdots,{\bf r}_{l-1};{\bf r}_l) &\equiv&
\langle \psi^\dagger_g({\bf r}_1)\cdots\psi^\dagger_g({\bf r}_{l-1})
{\bf P}^+({\bf r}_l) \psi_g({\bf r}_{l-1})\cdots
\psi_g({\bf r}_1)\rangle,\\
\rho_l({\bf r}_1,\cdots,{\bf r}_l) &\equiv& \langle
\psi_g^\dagger({\bf r}_1)\cdots
\psi_g^\dagger({\bf r}_l) \psi_g ({\bf r}_l)\cdots \psi_g ({\bf r}_1)
\rangle,
\eea
\label{eq:expect}
\end{mathletters}
and considering a $J_g=0\rightarrow J_e=1$ transition for simplicity, we
were able to cast the hierarchy of equations for atomic correlations
functions into the following form {\cite{RUO97a}}
\bea
\lefteqn{\dot{\bf P}_l({\bf r}_1,\cdots,\rv_{l-1};{\bf r}_l)=
(i\delta-\gamma) {\bf P}_l({\bf r}_1,\cdots,\rv_{l-1};{\bf
r}_l)}\nonumber\\ &&+\sum_{k=1}^{l-1}{\sf G}({\bf r}_l-{\bf r}_k) {\bf
P}_l({\bf r}_1,\cdots,\rv_{k-1},\rv_{k+1},\cdots,\rv_l;{\bf r}_k)
+i\kappa \rho_l({\bf r}_1,\cdots,{\bf r}_l) {\bf D}^+_F({\bf
r}_l)\nonumber\\ &&+\int d^3r_{l+1}\,{\sf G}({\bf r}_l-{\bf r}_{l+1})
{\bf P}_{l+1}({\bf r}_1,\cdots,{\bf r}_l;{\bf r}_{l+1})\,.
\label{eq:hie}
\eea
The quantity ${\bf P}_l$ reflects correlations between the dipole moment of one
atom and the positions of $l-1$ other atoms, and $\rho_l$ is simply the density
correlation function for $l$ ground state atoms.

In Eq.~{(\ref{eq:hie})} the atoms are regarded as point dipoles. As we
have already emphasized in Ref. \cite{RUO97a}, this assumption must fail
at short distances. Real atomic interaction potentials are thought to
have a hard core, which prevents the atoms from overlapping.
Mathematically, one expects that all atomic correlation functions such as
${\bf P}_l$ and $\rho_l$ vanish if any two position arguments are the
same. Evidently, the contact interaction should be omitted as
inconsequential in Eq.~{(\ref{eq:hie})}. In view of
Eq.~{(\ref{eq:DOL})}, this is done by removing the delta function
contributions from the field propagator
${\sf G}$ by the substitution ${\sf G}_{ij}\rightarrow{\sf G}_{ij}'$,
with
\begin{equation}
{\sf G}_{ij}'(\rv)={\sf G}_{ij}(\rv)+i\kappa\delta_{ij}\delta(\rv)/3\,.
\end{equation}

The key mathematical insight of this paper is that the same substitution
applies even for point dipoles. In other words, the resulting
light-matter dynamics is exactly the same, whether we use the field
propagator ${\sf G}$ or ${\sf G}'$ in Eq.~{(\ref{eq:hie})}. To demonstrate
this contention, we consider the steady-state solutions of
Eqs.~{(\ref{eq:hie})}. For each $l$ we have a system of $l$ coupled equations
for $\Pv_l(\cdots;\rv_1)$, $\Pv_l(\cdots;\rv_2)$,...,
$\Pv_l(\cdots;\rv_l)$. The
coupling arises from the collective line shifts and linewidths, those ${\sf G}$
terms in Eq.~{(\ref{eq:hie})} that appear outside the integral. The idea is
that, if we solve this system of equations for a fixed $l$, in
$\Pv_l(\cdots;\rv_l)$ and indeed even in the seemingly more divergent
expression
${\sf G}(\rv_k-\rv_l)\,\Pv_l(\cdots;\rv_l)$ with $k\neq l$ the delta functions
of the form $\delta(\rv_j-\rv_s)$ $(j\neq s)$ cancel. The $\delta$
functions are inconsequential in integrals of the type $\int d^3r_{l-1}\,{\sf
G}(\rv_{l-1}-\rv_l)\,\Pv_l(\cdots;\rv_l)$, and hence in the solution of the
hierarchy (\ref{eq:hie}) for $\Pv_1$.

We begin the detailed discussion by noting that in the derivation of
Eqs.~{(\ref{eq:MonoD})} and {(\ref{eq:hie})} in Ref.~\cite{RUO97a} we used a
cutoff in the wave numbers $q$ of the photons. Here we again invoke the 
high-frequency cutoff. The $\delta$ function is then regarded an ordinary
function, albeit one that is sharply peaked over a small distance scale
$\alpha$. Seemingly unorthodox operations such as divisions by a $\delta$
function are therefore well defined. After the calculations we set
$\alpha\rightarrow0$.

The steady-state solutions for $\Pv_l(\cdots;\rv_k)$, $k=1,\cdots,l$, of
Eq.~{(\ref{eq:hie})} are obtained from a system of linear equations
$A{\bf x}={\bf b}$, where the $l\times l$ matrix $A$ and the $l\times1$
vectors ${\bf x}$ and ${\bf b}$ are defined by
\bea
A_{kq}&=&\delta_{kq} - (1-\delta_{kq}) {\alpha\over i\kappa} {\sf
G}(\rv_k-\rv_q);\quad x_k=\Pv_l(\cdots;\rv_k);\nonumber\\
b_k&=&\alpha
\rho_l({\bf r}_1,\cdots,{\bf r}_l) {\bf D}^+_F({\bf r}_k)+{\alpha\over
i\kappa}\int d^3r_{l+1}\, {\sf G}({\bf r}_k-{\bf r}_{l+1}) {\bf
P}_{l+1}({\bf r}_1,\cdots,{\bf r}_l; {\bf r}_{l+1})\,.\nonumber
\eea
Here $\alpha=-\kappa/(\delta+i\gamma)$ is the polarizability of a single
atom. The solution for this linear system may be expressed using
Cramer's rule \cite{CRA}. In particular,
$x_l=\Pv_l(\cdots;\rv_l)=\det{[\bar{A}^{(l)}]}/\det{(A)}$, where
$\bar{A}^{(l)}_{kq}=A_{kq}$, for $q\neq l$, and $\bar{A}^{(l)}_{kl}=b_l$. For 
any two indices $j\neq s$, the matrix $A$ contains two entries 
${\sf G}(\rv_j-\rv_s)$,
one on the row $j$ and column $s$ and the other on row $s$, column $j$. The
determinant $\det{(A)}$ therefore is a quadratic polynomial of the
propagator ${\sf G}(\rv_j-\rv_s)$. The coefficient of ${\sf G}(\rv_j-\rv_s)^2$
is a function of the position variables $\{\rv_i\}$, and may
technically be zero for some sets of coordinates $\{\rv_i\}$. In our
argument we assume that such accidentally zeros, if any, are inconsequential
in the physics. We thus posit that $\det{(A)}$ contains a term proportional to
${\sf G}(\rv_j-\rv_s)^2$, and hence is a proper quadratic polynomial of
$\delta(\rv_j-\rv_s)$. By the same token,  $\det{[\bar{A}^{(l)}]}$ is a second
order polynomial of $\delta(\rv_j-\rv_s)$ for $j\ne s$, $j\ne l$, $s\ne l$, and
a {\em first\/} order polynomial in $\delta(\rv_j-\rv_l)$, $j\ne l$.

Let us first consider the case $l=N$, where $N$ is the total number of atoms.
All the expectation values in Eq.~{(\ref{eq:expect})} vanish for $l>N$. Thus,
$\Pv_{N+1}=0$ and $b_j=\alpha\rho_N\Dv^+_F(\rv_j)$, for $j=1,\cdots,N$. By the
preceding argument, $\Pv_{N} = \det[\bar A^{(N)}]/\det(A)$ is a rational
expression whose denominator is a quadratic polynomial in each
$\delta(\rv_j-\rv_s)$,
$j\ne s$, whereas the numerator is a quadratic expression of
$\delta(\rv_j-\rv_s)$ with
$j\ne s$, $j\ne N$, $s\ne N$, and a first-order polynomial in
$\delta(\rv_j-\rv_N)$, $j\ne N$. In the expression ${\sf
G}(\rv_N-\rv_{N-1})P_N(\rv_1,\cdots,\rv_{N-1};\rv_N)$ all the delta function
divergences in the numerator are therefore canceled by equal (or higher) powers
of the same delta functions in the denominator.

Next consider the case $l=N-1$. Because the expression ${\sf
G}(\rv_N-\rv_{N-1})P_N(\ldots;\rv_N)$ does not contain
uncanceled delta functions, its integral over $\rv_N$  needed to calculate the
$(N-1)\times 1$ vector ${\bf b}$ for $l=N-1$ does not contain any delta
function divergences either. With similar arguments as before we then conclude
that also ${\sf G}(\rv_{N-2}-\rv_{N-1})\,\Pv_{N-1}(\cdots;\rv_{N-1})$ is free
from uncanceled delta functions. By repeating the procedure for all
$l$ we confirm that the expressions ${\sf
G}(\rv_{l-1}-\rv_l)\,\Pv_l(\cdots;\rv_l)$, for any $l$, do not contain
delta functions which could affect the integrations.

Because our derivation is very technical, we give a simple example for the
case of two atoms. Then, $\Pv_3=0$. The ground state density and
the polarization at $\rv_1$ and $\rv_2$ are coupled by the collective
linewidths and line shifts leading to the resonant dipole-dipole interaction
\cite{RUO97a}. For $l=2$, $A_{11}=A_{22}=1$,
$A_{12}=A_{21}=-\alpha{\sf G}(\rv_1-\rv_2)/i\kappa$, and $\Pv_2(\rv_1;\rv_2)$
may be expressed using Cramer's rule. In particular,
\bea
\lefteqn{\int d^3 r_2\, {\sf G}(\rv_1-\rv_2)\,\Pv_2(\rv_1;\rv_2)}\nonumber\\
&&=\int d^3 r_2\, {\sf
G}(\rv_1-\rv_2)\,\alpha\rho_2(\rv_1,\rv_2){\Dv_F^+(\rv_2)
+\alpha{\sf G}(\rv_1-\rv_2)\Dv_F^+(\rv_1)/i\kappa\over 1-
(\alpha{\sf G}(\rv_1-\rv_2)/i\kappa)^2}\,.
\label{eq:P2}
\eea
With $\rv_2\neq\rv_1$, the delta functions in ${\sf G}(\rv_1-\rv_2)$ vanish,
and at $\rv_2=\rv_1$ we may cancel $[\delta(\rv_1-\rv_2)]^2$. The delta
functions do not affect the integral because the highest power of
$\delta(\rv_1-\rv_2)$ is two, in both the denominator and the numerator. Thus,
the contact interactions do not affect the polarization $\Pv_1(\rv_1)$,
which is obtained by inserting the expression {(\ref{eq:P2})} into
Eq.~{(\ref{eq:hie})}.

We have shown that all delta functions contained in $\sf G$ cancel in
Eq.~{(\ref{eq:hie})}. A corollary of the derivation is that the polarization
self-energy term (\ref{eq:pse}) does not have any effect on the light-matter
dynamics for point dipoles, and may be ignored.

There is a concurring physical explanation for our result.
Comparison of Eqs.~(\ref{eq:DOL}) and~(\ref{eq:hie}) shows that the
delta function lumps with the detuning $\delta$, and gives a divergent
frequency shift. As the dipole at ${\bf r}_k$ draws nearer to the
dipole at ${\bf r}_q$ and the electric fields of the dipoles on each
other grow stronger, at the same time the dipoles are also shifted
further and further away from resonance with one another. Within our
present viewpoint it is the level shift due to resonant dipole-dipole
interactions that dynamically curbs the interaction energy between the
dipoles.

After arguing that the delta functions may be ignored in
Eq.~{(\ref{eq:hie})}, we now rewrite the example presented in
Ref.~{\cite{RUO97a}} for the optical response of a homogeneous
condensate with the contact interactions omitted. The condensate is
assumed to fill the half-space $z\geq0$ with the constant density
$\rho$; for an ideal condensate the density $\rho$ immediately implies the
other correlation functions as $\rho_l = \rho^l$. The optical response
is solved in an approximation that ignores the collective line shifts
and linewidths. Specifically, we replace ${\sf G}$ in our earlier
argument by ${\sf G}'$, and then ignore all tensors ${\sf G}'$ that
reside outside integrals in Eq.~{(\ref{eq:hie})}. This approximation is only
guaranteed to be valid for dilute condensates, $\rho\lambda^3\ll 1$, where
$\lambda$ is the wavelength of light \cite{com1}. What the exact limit
of validity is and what types of corrections emerge at high densities
is at the moment largely unknown.

We consider the steady-state solution of Eqs.~{(\ref{eq:hie})} with
$l=1,\ldots$. The initial free field is written $\Dv^+_F(\rv)=D_F\, \pol
\exp{(ikz)}$. The following set of damped plane waves solves
Eqs.~{(\ref{eq:hie})}:
\begin{equation}
{\bf P}_l({\bf r}_1,\ldots;{\bf r}_l) =
\left\{
\begin{array}{ll}P\,\rho^{l-1}\,\hat{\bf e}\,e^{ik'z_l},&z_1,
\ldots z_l\ge0;\\ 0, &{\rm otherwise}\,,
\end{array}
\right.
\label{eq:NCS}
\end{equation}
with $k'$ ($\Im(k') >0$) and $P$ being the variable parameters. This may be
seen
by evaluating the following integral for $z_1>0$
\bea
\lefteqn{\int_{z_2\ge0}d^3r_2\,\hat{\bf e}^* \cdot{\sf
G}'({\bf r}_1-{\bf r}_2) \cdot\hat{\bf e}\, e^{ik'z_2}}\nonumber\\
&&= i\kappa\left[{2k^2+k'^2\over 3( k'^2-k^2)}\,e^{ik'z_1} +
{{k'}^2\over2k(k-k')}\,e^{ikz_1}\right]\,.
\label{eq:PRI}
\eea
The integrals on the right-hand side of (\ref{eq:hie}) produce
sums of two exponentials, the vacuum component $\propto e^{ikz}$, which
should cancel the free-field terms, and a remaining $e^{ik'z}$ term,
which pairs up with the correlation functions ${\bf P}_l$. All of
the Eqs.~(\ref{eq:hie}) then reduce to the following two conditions:
\begin{mathletters}\label{eq:SSC}
\begin{eqnarray}
\left(i\delta-\gamma + {i\rho\kappa(2 k^2+k'^2)\over 3(k'^2-k^2)} \right)P
&=& 0
\,,\label{eq:SS1}\\
D_F + {{k'}^2\over 2k(k-k')}\,P &=& 0\,,
\label{eq:SS2}
\end{eqnarray}
\end{mathletters}
which give the wave number $k'$ and the polarization amplitude $P$.

On the other hand, Eq.~{(\ref{eq:MonoD})} for the electric field refers to a
one-atom quantity $\Pv^+\equiv\Pv_1$. There is no evident reason to drop the
delta function contribution in ${\sf G}$, so we use Eq.~{(\ref{eq:MonoD})}
as it
is. The total electric field is assumed to be of the form $E\,\hat{\bf
e}\,e^{ik'z}$. With the choice (\ref{eq:SS2}), the vacuum type
contributions $e^{ikz}$ indeed cancel, and the polarization amplitude is
related
to the amplitude of electric field by
$
\epsilon_0 E = k^2P/( k'^2-k^2).
$
With the help of Eq.~{(\ref{eq:SS1})} we find
\begin{equation}
n^2 - 1  = {\rho\alpha\over(1-\rho\alpha/3)}\,,
\label{eq:CHI}
\end{equation}
where the refractive index $n$ is defined in a familiar manner, $k' = nk$.
Because we are dealing with a linear theory, the electric field and the
polarization are naturally related by $\Pv^+ = \epsilon_0\chi \Ev^+$; but
in addition we now find the familiar relation $\chi = n^2-1$ between
susceptibility and refractive index.

It should be noted that keeping or ignoring the delta function of $\sf G$
in Eq.~{(\ref{eq:MonoD})} simply makes a difference between two different
definitions of electric field inside a dielectric medium. This is a thorny
and often an inconsequential issue. The final measurements are usually carried
out outside the dielectric, whereupon the delta function does not contribute
anyway. Our choice has the advantage that it aligns with the standard
conventions of electrodynamics.

While we earlier \cite{RUO97a} obtained the column density result,
$n^2-1=\chi = \rho\alpha$, Eq.~(\ref{eq:CHI}) is the classic column density
result with the added LL local-field correction \cite{LOR}. As far as our
microscopic argument is concerned, the correction emerges as a result of
divergent level shifts of the atoms due to dipole-dipole interactions.
Within the present approach the main approximations for the LL shift are
two: collective linewidths and line shifts are ignored, and correlation
functions for ground state atoms as appropriate for a BEC are assumed.
Atom-atom correlation functions, though trivial, are formally taken into
account to infinite order. In comparison, Ref.~\cite{MOR95} strives at
taking into account collective linewidths and line shifts, but in exchange
makes approximations concerning atomic correlation functions.

Apart from heralding the physics insights that originally lead to
the Lorentz-Lorenz formula, our results should have practical consequences
in the ongoing studies of the condensates. The LL shift could, and
possibly should, be studied as the first correction to the column density
arguments, which so far have been the exclusive tool in the analysis of
the experimental results. As another example, we conjectured in
\cite{RUO97a} that the entire linear hierarchy of Eqs.~(\ref{eq:hie}) for
$l=1,2,\ldots,N$ may be solved by solving numerically the classical equations
for the coupled system of electromagnetic fields and $N$ charged harmonic
oscillators. The position correlation functions $\rho_l$ could be taken into
account by repeating the solution for a number of initial configurations chosen
from the proper stochastic ensemble, and averaging the results. This type of a
Monte-Carlo approach would seem difficult in the presence of delta functions,
because they dwell in a set of measure zero in phase space of the positions of
the atoms and yet may substantially affect the results. However, as the
troubling delta functions have now proven inconsequential, the door is ajar for
an essentially exact numerical solution of the optical response of a dense,
near-resonance BEC.

We would like to thank Robert Graham and Craig Savage for comments on the
manuscript. This work is supported in part by the National Science Foundation,
grant number PHY-9421116.


\begin{references}
\bibitem{LOR} In some contexts the
local-field correction to electric susceptibility is also known as the
Clausius-Mossotti relation.
\bibitem{com1} This may be shown to correspond to the expansion in which
only all the scattering processes by isolated atoms are included,
and scattering processes in which a photon undergoes multiple scatterings
between pairs, triplets, etc., of atoms are ignored.
\end{references}
\end{document}